\documentclass{article}

\usepackage{arxiv}

\usepackage[utf8]{inputenc} 
\usepackage[T1]{fontenc}    
\usepackage{hyperref}       
\usepackage{url}            
\usepackage{booktabs}       
\usepackage{amsfonts}       
\usepackage{nicefrac}       
\usepackage{microtype}      
\usepackage{lipsum}
\usepackage{graphicx}
\graphicspath{ {./images/} }

\usepackage[numbers]{natbib}  
\usepackage{amsmath} 
\usepackage{float} 
\usepackage{placeins} 

\title{Hybrid Branch-and-Price with Quantum-Inspired Pricing for Intra-Day Electric Vehicle Charging Scheduling via Partition Coloring}

\author{
 Peng Sun \\
  College of Management and Economics\\
  Tianjin University\\
  Tianjin, 300072, China \\
  \texttt{sunpeng1317@tju.edu.cn} \\
   \And
 Liang Zhong \\
  College of Management and Economics\\
  Tianjin University\\
  Tianjin, 300072, China \\
  \texttt{zl-2002@tju.edu.cn} \\
  \And
 Qing-Guo Zeng \\
  Shenzhen Institute for Quantum Science and Engineering\\
  Southern University of Science and Technology\\
  Shenzhen, 518055, China \\
  \texttt{zengqg2022@mail.sustech.edu.cn} \\
   \And
  Li Wang\thanks{Corresponding Author}\\
  College of Economics and Management\\
  Hebei University of Technology\\
  Tianjin, 300130, China \\
  \texttt{li.wang@hebut.edu.cn} \\
}

\begin{document}
\maketitle
\begin{abstract}
The rapid deployment of electric vehicles (EVs) in public parking facilities and fleet operations raises challenging intraday charging scheduling problems under tight charger capacity and limited dwell times. We model this problem as a variant of the Partition Coloring Problem (PCP), where each vehicle defines a partition, its candidate charging intervals are vertices, and conflicts represented as edges include temporal overlap, resource incompatibility, and same-partition exclusivity. On this basis, we design a branch-and-price algorithm in which the restricted master problem selects feasible combinations of intervals, and the pricing subproblem is a maximum independent set problem. The latter is reformulated as a quadratic unconstrained binary optimization (QUBO) model and solved by quantum-inspired algorithms (QAIA) implemented in the MindQuantum framework, specifically the ballistic simulated branching (bSB) and simulated coherent Ising machine (SimCIM) methods, while the master problem is solved by Gurobi. Both bSB and SimCIM are executed on classical hardware; thus, the proposed approach is a classical decomposition framework enhanced by quantum-inspired heuristics, rather than a quantum hardware algorithm. Computational experiments on a family of synthetic EV charging instances show that the QAIA enhanced algorithms match the pure Gurobi baseline branch-and-price method on small and medium instances, and clearly outperform it on large and hard instances. In several cases where the baseline reaches the time limit with nonzero optimality gaps, the QAIA variants close the gap and prove optimality within the same time budget. These results indicate that integrating QAIA into classical decomposition schemes is a promising direction for large-scale EV charging scheduling and related PCP applications.
\end{abstract}


\section{Introduction}
In recent years, the widespread adoption of electric vehicles (EVs) in public parking facilities, workplace charging stations, and large fleet operations has made intraday charging scheduling an important operational optimization problem. Unlike conventional refueling, EV charging is time-consuming and highly flexible; it is influenced by users’ departure deadlines, power capacity limitations, and fluctuations in instantaneous charging load. In typical intraday scenarios, vehicles arrive and depart within the same day and expect to obtain sufficient energy during a limited dwell period to support subsequent trips. Consequently, operators must assign feasible charging intervals to each vehicle under restricted resources while balancing demand satisfaction, charger utilization, and overall system stability.

The literature has examined the EV charging scheduling problem from multiple perspectives. For instance, Dolgui et al. \citep{dolgui2025scheduling} formulated a scheduling model for conventional charging tasks with parallel chargers and explicitly characterized task-resource conflicts, while Wang et al. \citep{wang2024joint} developed an integrated framework that jointly models daily mobility patterns and charging activities. In addition, Shen et al. \citep{shen2025coordinating} proposed a two-stage method that coordinates vehicle-charger matching and bidirectional charging strategies under photovoltaic uncertainty. Exact optimization approaches, including mixed integer linear programming (MILP) and column generation methods, have been systematically developed; however, their scalability to large instances remains limited due to rapid growth in problem size and computational complexity. To handle large-scale and dynamic environments, data-driven and heuristic approaches, including deep learning \citep{chemudupaty2025optimizing}, reinforcement learning \citep{xu2025impact,korkas2024distributed}, and genetic algorithms \citep{wu2023electric}, have also been widely applied to derive near-optimal charging policies.

Most existing studies rely on time-indexed integer programming models. By discretizing vehicles’ dwell windows, scheduling decisions are mapped to binary assignments between vehicles and time slots. Although effective for small and medium instances, this modeling approach has inherent and documented limitations that persist despite recent algorithmic advances. The model size grows rapidly as the temporal resolution increases, rendering even advanced solvers ineffective when fine grained time discretization is needed; conflict relationships among candidate charging intervals, such as temporal overlap, sharing of physical chargers, or local load interactions, are only implicitly represented and difficult to exploit algorithmically; and the number of constraints and variables escalates quickly with the number of vehicles and candidate intervals, approaching intractability for realistic problem sizes. While column generation and cutting plane techniques have improved performance on some structured instances \citep{parmentier2023electric,zang2022column}, large-scale instances with rich resource coupling and conflicting constraints still present significant computational challenges that demand alternative modeling and solution paradigms.

Graph-based formulations offer a natural way to represent conflicts explicitly, particularly for capturing the mutual exclusivity of candidate charging intervals. Recent work has shown that the scheduling problem can be reformulated as a vertex coloring problem and solved using the Quantum Approximate Optimization Algorithm (QAOA) \citep{griset2024smartcharging}. However, EV scheduling exhibits a specific structural characteristic that pure vertex coloring does not capture: each vehicle may have several candidate intervals from which exactly one must be selected, reflecting hard disjunctive constraints inherent to vehicle scheduling. This “vehicle as partition, interval as vertex, exactly one vertex selected per partition” structure naturally aligns with the Partition Coloring Problem (PCP), which provides a more direct and constrained formulation. By explicitly representing the partition structure, the PCP enables a tighter mathematical characterization of the problem and potentially allows more sophisticated decomposition and branch and bound algorithms to exploit the special structure. Although the PCP has been studied in combinatorial optimization \citep{furini2018exact,zhu2020partition,cseker2022digital}, it has not been systematically applied to EV charging scheduling.

To address this gap, we model the intraday EV charging scheduling problem as a variant of PCP. Let $N$ be the set of vehicles, and let each vehicle $n \in N$ define one partition $P_n$. Candidate charging intervals are represented as vertices in $\mathcal{I}$, and edges in $\mathcal{E}^A$ connect conflicting intervals (temporal overlap, resource incompatibility, or belonging to the same vehicle partition). With a charger set $\mathcal{C}=\{1,\dots, C\}$, the objective is to select exactly one interval from each partition and construct a feasible schedule that minimizes the makespan.

Based on this modeling framework, our main contributions are as follows:

\begin{enumerate}
    \item We provide the first systematic formulation of the intraday EV charging scheduling problem as a PCP, enabling explicit representation of vehicle-level and interval-level conflicts while naturally supporting multiple resource constraints.
    \item We develop a Branch-and-Price algorithm in which the master problem handles partition selection and resource allocation, and the pricing subproblem generates new columns via a maximum independent set structure, thereby improving computational efficiency.
    \item We integrate QAIA solvers (bSB and CFC), implemented in the MindSpore Quantum framework and executed on classical hardware, to accelerate the pricing subproblem and enhance scalability for large problem instances.
\end{enumerate}

Finally, we conduct extensive numerical experiments comparing two QAIA-enhanced branch-and-price algorithms, bSB and CFC, with a pure Gurobi baseline on synthetic EV charging instances of varying sizes, demonstrating that these QAIA variants significantly improve performance on large and hard cases while preserving optimal solution quality.

The remainder of this paper is organized as follows. Section~\ref{sec:Related works} reviews the literature on EV charging scheduling and related optimization approaches. Section~\ref{sec:Problem} describes the problem. Section~\ref{sec:Methodology} presents the formal mathematical formulation. Section~\ref{sec:Experimental} reports the computational experiments and compares a pure Gurobi baseline branch-and-price method with two QAIA pricing variants, bSB and CFC, on a family of PCP instances derived from EV charging scenarios. Finally, Section~\ref{sec:Conclusions} concludes the paper and outlines potential directions for future research.

\section{Related Works}\label{sec:Related works}

\subsection{Electric Vehicle Charging Scheduling}
EV charging scheduling research has expanded rapidly, but the literature covers heterogeneous problem settings that should be distinguished before comparing algorithms. Recent reviews summarize this diversity from deterministic scheduling and machine-scheduling perspectives \citep{dolgui2025scheduling}, as well as reinforcement-learning-based charging control \citep{zhao2024reinforcement}. At the problem level, key dimensions include: offline vs. stochastic online decision-making, fixed/regular charging vs. flexible interval selection, preemptive vs. non-preemptive charging, and station-level power-envelope constraints.

Recent studies further enrich these problem settings. For example, offline and stochastic online formulations with renewable integration and request admission have been jointly studied in \citep{gauchotte2026study}. Stochastic charging-duration uncertainty with multi-objective performance criteria has been investigated in \citep{khiar2025multi}. Distribution-level power-envelope constraints and charger-health-aware fairness have also been explicitly modeled in \citep{huang2026health}. In addition, practical scenarios such as mobile-charging coordination and digital-twin-based rolling optimization have been explored in \citep{fang2026game,liu2026digital}. These works substantially improve realism and operational relevance in EV charging scheduling.

From a methods perspective, existing approaches include exact optimization, metaheuristics, and learning-based methods. Exact models remain the main tool for benchmarking and structural analysis \citep{dolgui2025scheduling,gauchotte2026study}, while stochastic and large-scale settings often rely on heuristics or learning to improve computational responsiveness \citep{khiar2025multi,zhao2024reinforcement}. Decomposition strategies such as column generation provide an effective middle ground for balancing solution quality and scalability \citep{parmentier2023electric,zang2022column}.

Despite this progress, two gaps remain for the intra-day setting considered here. First, at the problem level, many models emphasize energy cost, waiting, or grid-side objectives, but fewer studies focus on the structure in which each EV is associated with multiple contiguous feasible intervals, and exactly one interval must be selected under charger-capacity conflicts. Second, at the method level, there is still limited work that combines an exact decomposition framework with a pricing-oriented acceleration mechanism tailored to this partition-conflict structure. This paper addresses these two gaps by modeling the problem as a PCP and solving it with a branch-and-price framework enhanced by quantum-inspired pricing.

\subsection{Partition Coloring Problem}
The PCP was originally introduced in wavelength routing and assignment for optical networks, where exactly one vertex must be selected from each partition, and conflicting paths cannot share the same resource~\cite{li2000partition}. The problem is NP-complete, which explains the substantial computational difficulty observed in practical applications. From a modeling viewpoint, PCP extends the classical Vertex Coloring Problem (VCP): in addition to assigning colors under adjacency constraints, one must satisfy partition-selection constraints before coloring the induced subgraph. This extra combinatorial layer creates stronger coupling between feasibility and coloring decisions and typically requires dedicated algorithmic treatment.

Algorithmic studies on PCP span exact, heuristic, and hardware-accelerated directions. On the exact side, decomposition and cutting-plane frameworks improve performance on small and medium instances \citep{cseker2019decomposition, SEKER202167}, while set-packing-based formulations combined with Branch-and-Price provide stronger scalability and solution quality \citep{olariu2025set}. For larger instances, metaheuristics such as ant colony optimization offer robust approximate performance \citep{fidanova2014ant}. In parallel, digital-annealer-based QUBO approaches have been explored as alternative computational paradigms \citep{cseker2022digital}. Theoretical work further connects PCP to selective coloring, weighted coloring, and independent-set variants \citep{zhu2020partition}, broadening both its structural understanding and algorithmic transferability.

These developments make PCP a suitable abstraction for conflict-driven resource-allocation problems beyond optical networks, including spectrum assignment and EV charging scheduling \citep{kheiri2021constructing}. For the EV setting considered in this paper, the partition structure directly maps to the one-interval-per-vehicle requirement, while graph conflicts encode temporal and resource incompatibilities. This correspondence motivates adopting PCP as the core combinatorial model and integrating decomposition-based optimization with QAIA-enhanced pricing in our framework.

\subsection{Quantum Annealing Inspired Algorithm}

Many combinatorial optimization problems can be mapped to Ising models, where binary decisions are represented by spins $\sigma_i \in \{\pm1\}$ with pairwise couplings $J_{ij}$ and local fields $h_i$. The corresponding energy function is
\[
H = \frac{1}{2} \sum_{i,j} J_{ij} \sigma_i \sigma_j + \sum_i h_i \sigma_i .
\]
Computing the global ground state is NP-hard in general \cite{lucas2014ising}, which motivates physics-inspired optimization strategies that trade exactness for high-quality solutions and speed on hard instances.

This motivation has led to two closely related development paths. The first path is hardware-oriented, including coherent Ising machines (CIM), digital annealers, and related non-von-Neumann architectures \cite{mcmahon2016fully,aramon2019physics,cai2020power,borders2019integer}. The second path is algorithmic, where quantum-annealing-inspired algorithms (QAIA) emulate these physical dynamics on classical hardware \cite{zeng2024performance}. The algorithmic path is particularly attractive in practice because it avoids specialized hardware constraints while preserving much of the dynamical search behavior.

Representative QAIA families include simulated bifurcation (SB) methods \cite{kanao2022simulated} and variants of simulated coherent Ising machines (CIM) \cite{tiunov2019annealing}. A common strategy is to relax discrete spins into continuous variables, evolve them under annealing or bifurcation dynamics, and then project back to binary decisions. Since continuous relaxation can introduce analog errors, various error-correction and stabilization mechanisms have been developed. To mitigate the adverse effects of spin relaxation, multiple CIM variants have been numerically implemented on classical computers, including CIM with chaotic amplitude control (CAC), chaotic feedback control (CFC), and separated feedback control (SFC). While all these algorithms simulate various CIM dynamics, they differ in their feedback and correction strategies, with each variant designed to address specific challenges such as amplitude heterogeneity and convergence acceleration on large or hard instances. 

Among SB variants, ballistic simulated bifurcation (bSB) introduces inelastic boundary walls at $x_i = \pm 1$ to improve convergence robustness. The equations of motion for bSB are:
\begin{align}
\frac{dx_i}{dt} &= a_0 y_i \label{eq:bsb_x} \\
\frac{dy_i}{dt} &= -(a_0 - a(t)) x_i + c_0 \sum_{j=1}^{N} J_j x_j \label{eq:bsb_y}
\end{align}
where $x_i$ and $y_i$ denote position and momentum of the $i$-th oscillator, $a_0$ is a positive detuning frequency, $a(t)$ is a time-dependent pumping amplitude increasing from zero, and $c_0$ denotes the coupling strength of the Ising problem coefficients $J_j$. The inelastic walls force positions to be exactly $\pm 1$ when $a(t)$ becomes sufficiently large by setting $x_i := \text{sgn}(x_i) = \pm 1$ and $y_i := 0$ at each iteration when $|x_i| > 1$. This boundary control mechanism accelerates convergence by creating local minima at $[\pm 1, \pm 1]$ instead of relying purely on the nonlinear potential, thereby reducing continuous relaxation errors and improving the quality of projected binary solutions.

Among CIM variants, CFC is a refined algorithm designed to overcome amplitude heterogeneity and improve convergence on large instances. In the simplified CIM formulation, where nonlinear terms and imaginary parts of amplitude are dropped for computational efficiency, the basic evolution is:
\begin{align}
\frac{dx_i}{dt} = \left(vx_i + \zeta \sum_j J_j x_j\right) + \sigma f_i \label{eq:simcim}
\end{align}
where $\zeta$ is the coupling strength, $v$ denotes parametric gain, $\sigma$ represents linear loss coefficients, and $f_i$ is Gaussian noise. To address amplitude heterogeneity in this basic formulation, CFC introduces an auxiliary error variable $e_i$ controlled by feedback. The CFC method evolves the spin variable and error variable as follows:
\begin{align}
z_i &= -e_i \sum_j \zeta J_j x_j \label{eq:cfc_z} \\
\frac{dx_i}{dt} &= -x_i^3 + (p-1) x_i - z_i \label{eq:cfc_x} \\
\frac{de_i}{dt} &= -\beta e_i (z_i^2 - \alpha) \label{eq:cfc_e}
\end{align}
where $p$ and $\alpha$ are gain parameters, $\beta$ controls the rate of change of error variables, and $z_i$ is a feedback signal derived from error magnitude. The chaotic dynamics generated by this error-feedback mechanism enable more effective exploration of the search space near the ground state, thereby accelerating convergence and improving robustness on large instances.

In this work, we use bSB and CFC as QAIA engines for the pricing subproblem. bSB strengthens robustness by using inelastic boundary handling in bifurcation dynamics to accelerate convergence and jump out of local minima, while CFC employs chaotic-feedback error control to explore solution space more effectively and improve solution quality. Integrating these complementary QAIA solvers into the branch-and-price loop enables faster generation of high-quality pricing solutions for tightly coupled large-scale instances, while preserving the exact master-problem control in the hybrid framework.

\section{Problem Description}\label{sec:Problem}
\begin{figure}[h!]
\centering
\includegraphics[width=0.95\textwidth]{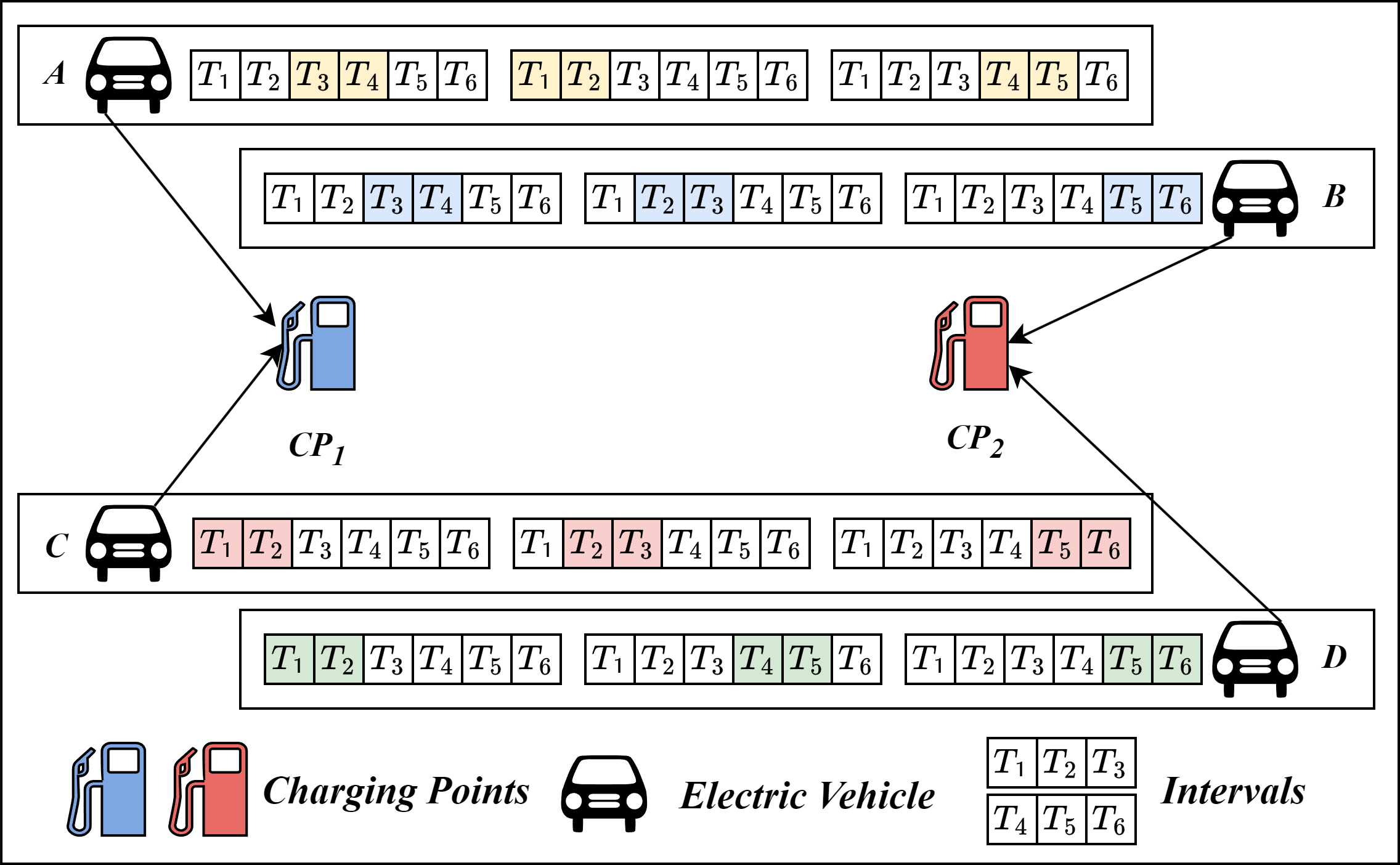}
\caption{Schematic illustration of the EV charging system. The station contains two chargers, and four EVs are shown, each associated with three candidate intervals. }
\label{fig:f0}
\end{figure}

This study considers a day-ahead charging scheduling problem for a fleet of EVs. Let $N$ denote the set of vehicles and $\mathcal{C}=\{1,\dots, C\}$ the set of identical chargers. The system configuration is illustrated in Fig.~\ref{fig:f0}. Each vehicle $n \in N$ arrives at a known time, and a set of feasible charging intervals is generated in advance from its arrival time, departure deadline, and energy requirement. For tractability, we assume that each candidate interval is one contiguous charging block that fully satisfies the demand of vehicle $n$, so interrupted or split charging is not allowed. As shown in Fig.~\ref{fig:f0}, four vehicles are depicted, each associated with three candidate intervals, and each vehicle must select exactly one interval from its corresponding partition.

\begin{figure}[h!]
\centering
\includegraphics[width=1\textwidth]{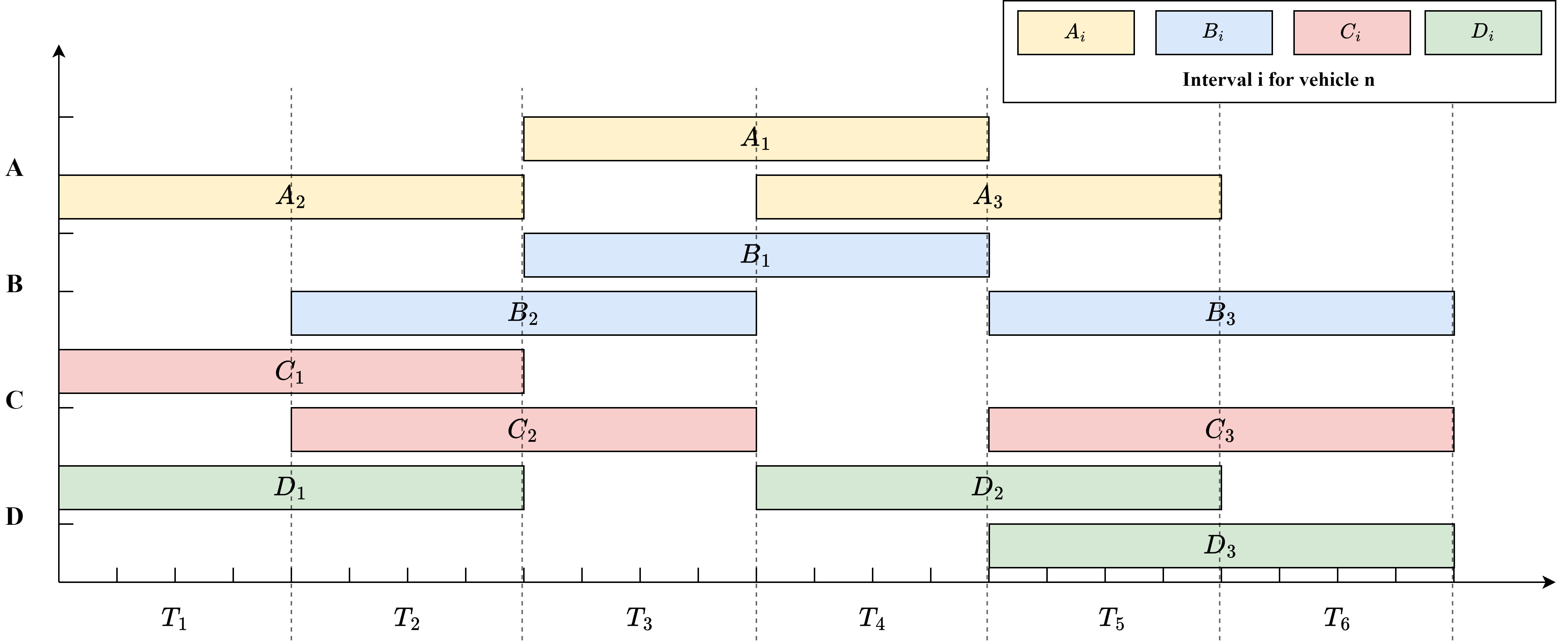}
\caption{Illustration of candidate charging intervals. The horizontal axis represents time, and the vertical axis lists vehicles $A$, $B$, $C$, and $D$. Each colored bar represents a candidate interval (e.g., $A_1$, $A_2$, $A_3$). Overlapping intervals across or within vehicles indicate conflicts, which form the edge set of the conflict graph used in the PCP formulation.}
\label{fig:f1}
\end{figure}

To describe the scheduling setting more concretely, we use the time-based representation shown in Fig.~\ref{fig:f1}. The horizontal axis represents time, and the vertical axis lists vehicles. Each colored bar corresponds to one candidate interval, and intervals of the same vehicle form one partition $P_n$. For example, for vehicle $A$, its three candidates $A_1$, $A_2$, and $A_3$ belong to the same partition and can be viewed as three vertices in that partition. Two intervals are in conflict if they belong to the same partition (i.e., correspond to the same vehicle) or if they belong to different partitions but overlap in time. A feasible schedule must satisfy: (i) exactly one interval is selected from each partition $P_n$; (ii) the number of simultaneously active intervals at any time does not exceed charger capacity $C$; and (iii) the makespan variable $\tau$ is minimized, where $\tau$ denotes the latest completion time among selected intervals and each interval completion time is denoted by $e_v$.

\begin{figure}[H]
    \centering
    \begin{minipage}{0.49\linewidth}
        \centering
        \includegraphics[width=\linewidth]{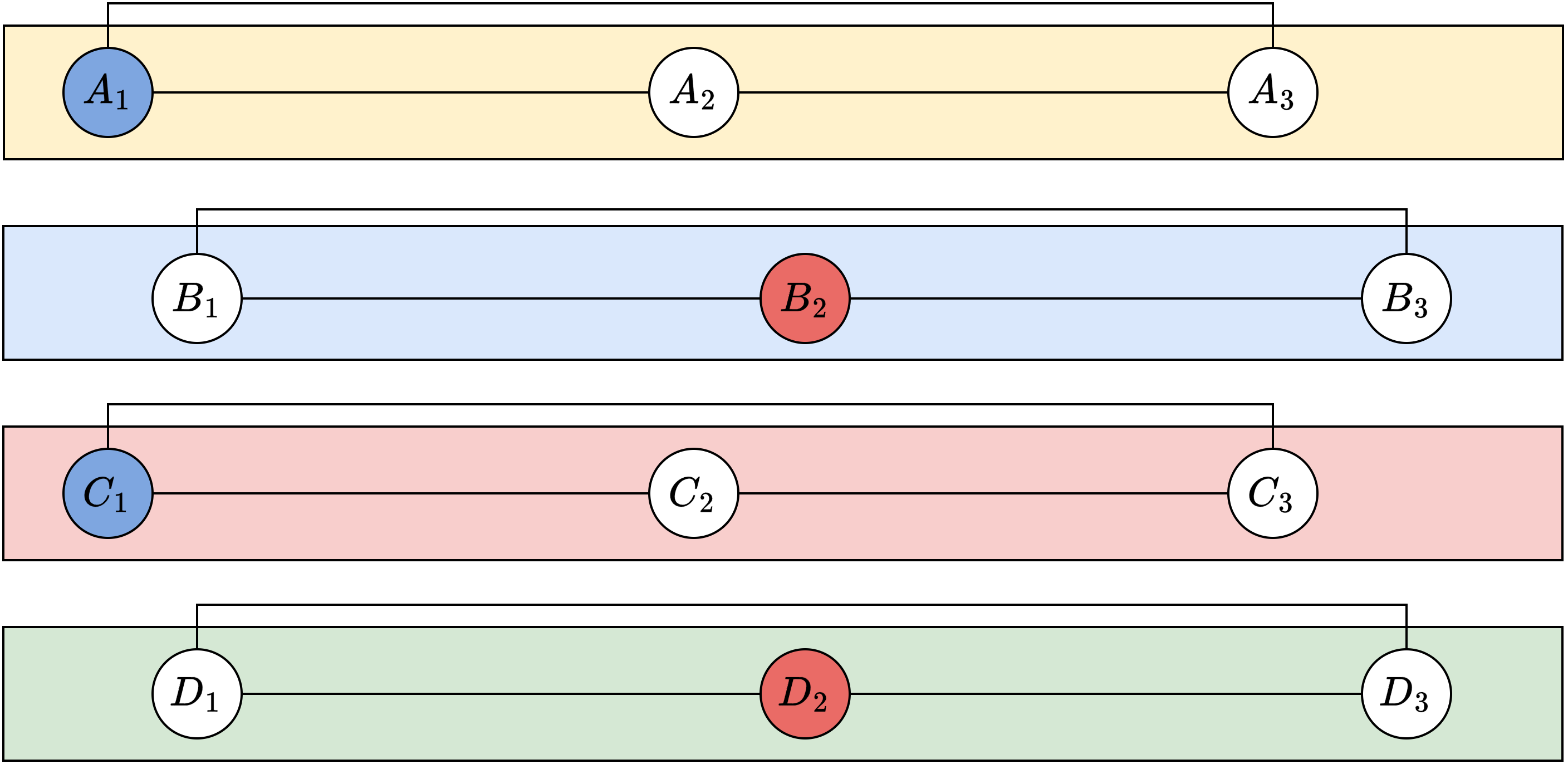} \\
        {\scriptsize (a)}
    \end{minipage}%
    \hfill
    \begin{minipage}{0.49\linewidth}
        \centering
        \includegraphics[width=\linewidth]{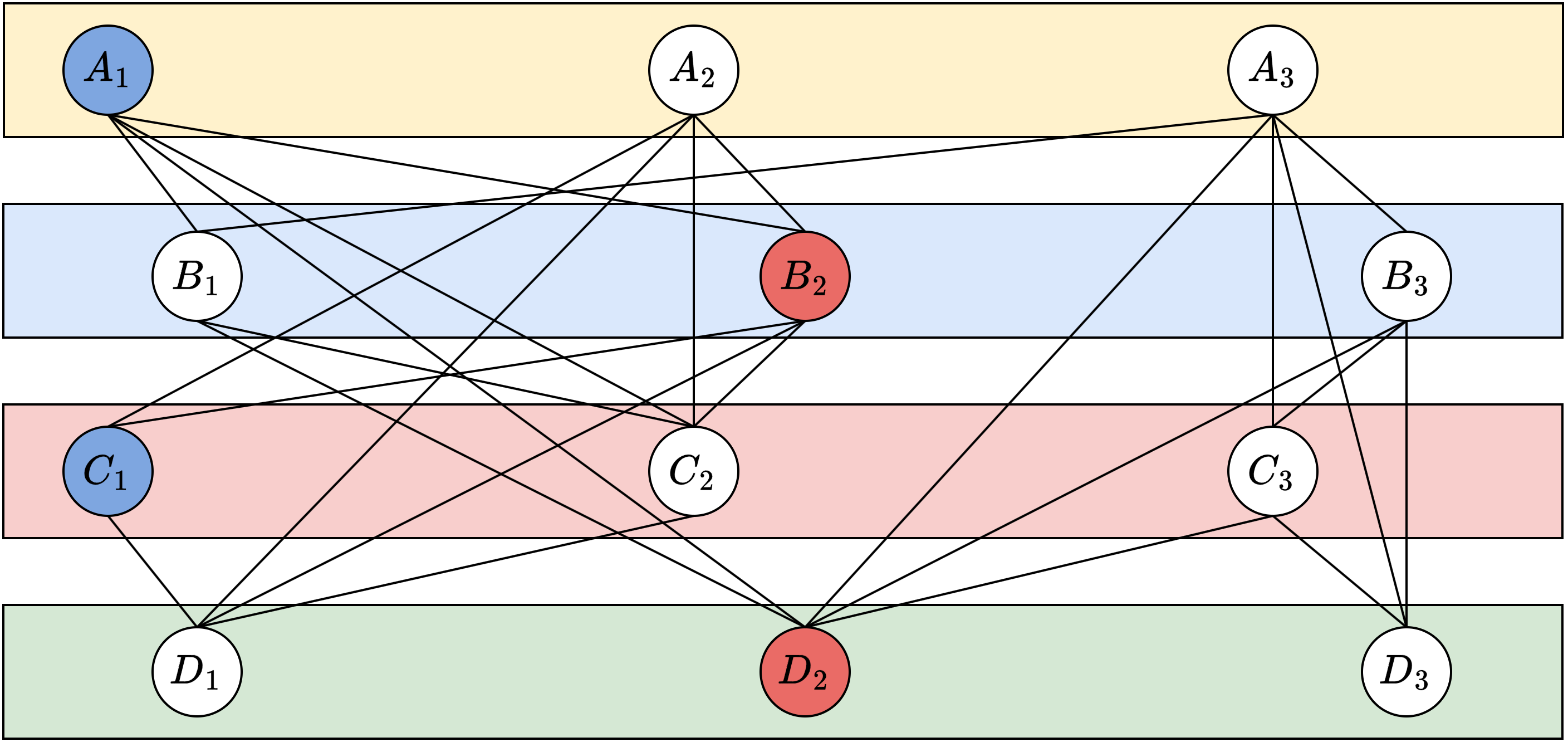} \\
        {\scriptsize (b)}
    \end{minipage}

    \vspace{0.5em} 

    \caption{Construction of the conflict graph for candidate charging intervals. 
    Subfigure~(a) illustrates intra-vehicle structure: for each vehicle (A, B, C, and D), the three candidate intervals (e.g., $A_1$, $A_2$, $A_3$) belong to one partition; in the auxiliary conflict graph, these same-vehicle vertices are pairwise connected (same-partition exclusivity), and thus induce a clique. 
    Subfigure (b) illustrates inter-vehicle conflicts: an edge connects two vertices from different partitions if their time windows overlap, meaning the two charging intervals cannot use the same charger. Vertices $A_1$ and $C_1$ (blue) form a compatible pair and can be allocated to one charger in the schedule. Likewise, $B_2$ and $D_2$ (red) are another compatible pair.}
    
\label{fig:f2}
\end{figure}

Figures~\ref{fig:f1} and~\ref{fig:f2} jointly illustrate the transformation from interval scheduling to the PCP graph model. Figure~\ref{fig:f1} shows candidate intervals on a timeline, while Figure~\ref{fig:f2} shows the corresponding graph structure: same-vehicle candidates belong to one partition and are connected by exclusivity edges, and cross-vehicle edges are added only for time-overlap conflicts.

To formalize the problem, let $G=(\mathcal{I},\mathcal{E}^A)$ be the auxiliary conflict graph induced by all candidate intervals. For each vehicle $n \in N$, its candidate vertices form a partition $P_n$, and the partition family is $\mathcal{P}=\{P_n \mid n \in N\}$. An edge $(u,v) \in \mathcal{E}^A$ exists if the corresponding intervals overlap in time or if $u$ and $v$ belong to the same partition, i.e., they are alternatives for one vehicle.

An independent set is a subset $S \subseteq \mathcal{I}$ such that no two vertices in $S$ are adjacent, i.e., $\forall u,v \in S$, $(u,v) \notin \mathcal{E}^A$. The charging-scheduling problem can then be expressed via PCP selection constraints together with charger assignment: selected vertices must satisfy one-per-partition feasibility, and compatible selected vertices can be grouped into charger-specific independent sets. In this interpretation, each charger $c \in \mathcal{C}$ corresponds to one independent-set group in the schedule.

A feasible PCP solution must satisfy the following conditions: (i) exactly one candidate charging interval is selected for each vehicle, i.e., $|S \cap P_n| = 1$ for all $n \in N$; and (ii) for any conflict edge $(u,v) \in \mathcal{E}^A$, vertices $u$ and $v$ cannot be assigned the same color; thus, each color class corresponds to a set of pairwise non-conflicting charging intervals that can be executed in parallel.

Under this formulation, each color class corresponds to a feasible subset of non-conflicting charging intervals that can be assigned to a single charging pile. Therefore, solving the PCP is equivalent to selecting independent sets that satisfy both the partition and conflict constraints. Each independent set selected in this way corresponds to a candidate column in the RMP, and the column generation procedure ensures that the resulting schedule is feasible.

\section{Methodology}\label{sec:Methodology}
\begin{figure}[h!]
\centering
\includegraphics[width=0.9\textwidth]{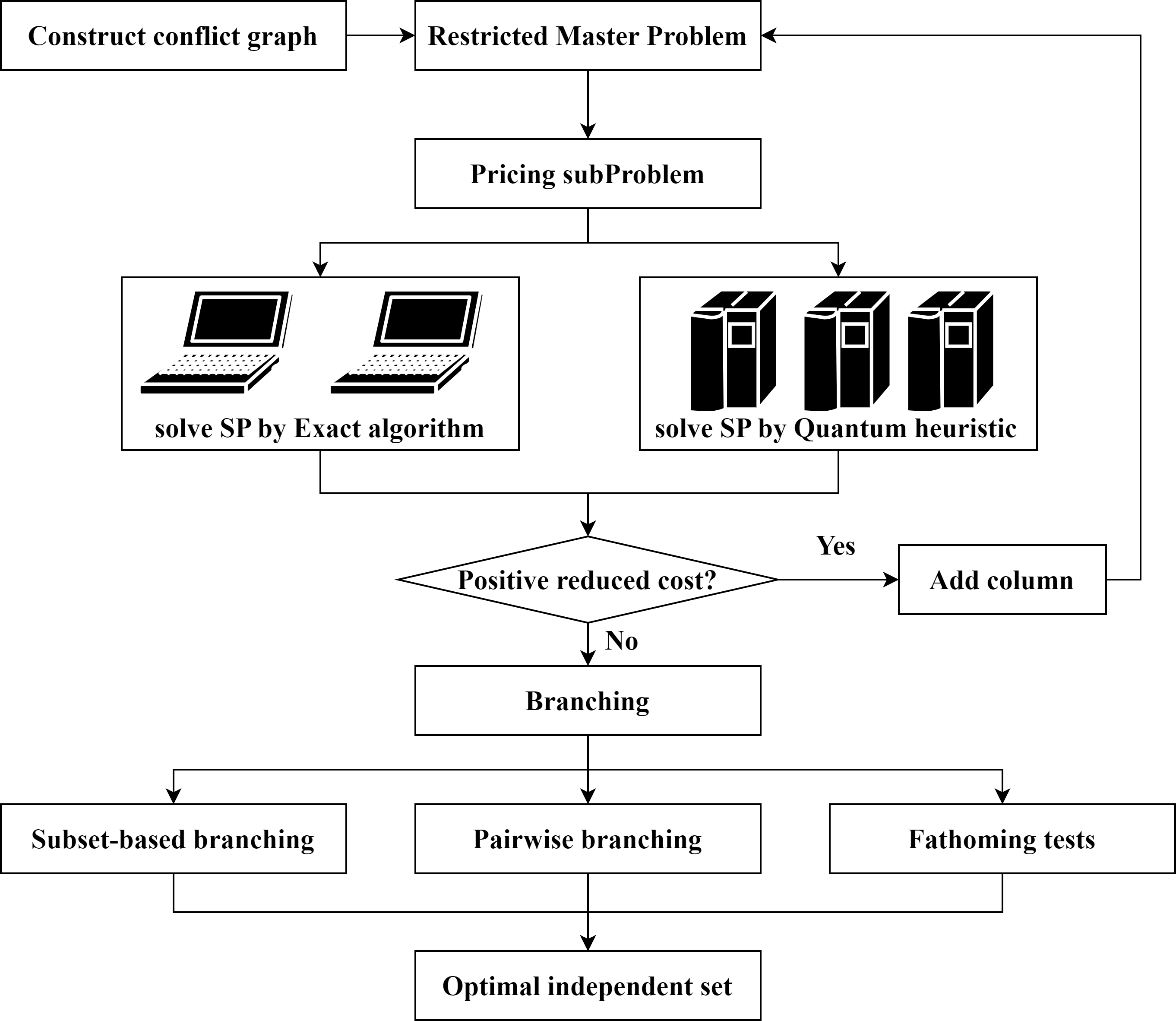}
\caption{Overview of the proposed solution framework for the EV charging scheduling problem.}
\label{fig:f4}
\end{figure}

Figure~\ref{fig:f4} summarizes the solution workflow. The method proceeds in four steps. We first generate feasible charging intervals for each EV and build a conflict graph based on temporal overlaps. We then solve the restricted master problem (RMP) in a column-generation framework and use the resulting dual information to identify promising new columns. Next, the pricing subproblem is written as a QUBO model and solved with QAIA; in parallel, we also solve the original pricing model with Gurobi as an exact benchmark. Finally, we apply a customized branch-and-bound procedure, designed around partition and pairwise-conflict structures, to obtain an integral and complete solution. This process handles interval conflicts, charger-capacity limits, and the one-interval-per-vehicle requirement in a unified way.

\subsection{ILP Formulations}

For clarity, we use $N$ to denote the set of vehicles and $\mathcal{I}$ to denote the set of candidate intervals, which are represented as vertices after the PCP transformation. We use $P_n$ for the subset of vertices belonging to vehicle $n$, and $e_v$ for the completion time of vertex $v$.

\begin{table}[!htbp]
\centering
\caption{Main symbols used in the model.}
\label{tab:notation}
\renewcommand{\arraystretch}{1.15}
\begin{tabular}{ll}
\toprule
Symbol & Meaning \\
\midrule
$N$ & Set of vehicles \\
$n$ & Vehicle index, $n \in N$ \\
$\mathcal{C}$ & Set of charging piles, $\mathcal{C}=\{1,\dots,C\}$ \\
$c$ & Charger index, $c \in \mathcal{C}$ \\
$\mathcal{I}$ & Set of all candidate intervals \\
$P_n$ & Partition of vehicle $n$ \\
$\mathcal{S}'$ & Current set of independent sets \\
$S$ & A candidate independent set \\
$\mathcal{E}^A$ & Conflict edge set of the auxiliary graph \\
$e_v$ & Completion time of vertex $v$ \\
$T_S$ & Completion time of column $S$, $T_S=\max_{v\in S} e_v$ \\
$\tau$ & Makespan variable in the RMP \\
$\zeta_{S,c}$ & RMP variable for column $S$ assigned to charger $c$ \\
$x_{v,c}$ & Binary pricing variable, $1$ if $v$ is selected on charger $c$ \\
$w_v$ & Vertex weight used in pricing objective \\
$\rho_1$ & Penalty coefficient for partition constraints in QUBO \\
$\rho_2$ & Penalty coefficient for conflict constraints in QUBO \\
$\mathcal{H}_0,\mathcal{H}_1,\mathcal{H}_2$ & QUBO objective term and penalty terms \\
$\mathcal{H}$ & Total QUBO Hamiltonian \\
\bottomrule
\end{tabular}
\end{table}

\noindent\textbf{Restricted Master Problem}

At each node of the Branch-and-Price algorithm, the RMP is solved over the current column set $\mathcal{S}'$:
\begin{equation}
\min_{\tau,\zeta}\ \tau
\label{eq:rmp-obj}
\end{equation}
subject to
\begin{equation}
\sum_{S\in\mathcal{S}':\, S\cap P_n\neq\emptyset}\ \sum_{c\in\mathcal{C}} \zeta_{S,c}=1,
\quad \forall n\in N,
\label{eq:rmp-partition}
\end{equation}
\begin{equation}
T_S \sum_{c\in\mathcal{C}} \zeta_{S,c}\le \tau,
\quad \forall S\in\mathcal{S}',
\label{eq:rmp-makespan}
\end{equation}
\begin{equation}
\sum_{S\in\mathcal{S}':\, u\in S \,\lor\, v\in S} \zeta_{S,c} \le 1,
\quad \forall (u,v)\in\mathcal{E}^A,\ \forall c\in\mathcal{C},
\label{eq:rmp-conflict}
\end{equation}
\begin{equation}
\zeta_{S,c}\ge 0,\ \forall S\in\mathcal{S}',\forall c\in\mathcal{C}, \qquad \tau \ge 0.
\label{eq:rmp-nonneg}
\end{equation}

Here,
\begin{equation}
T_S := \max_{v\in S} e_v,
\label{eq:rmp-ts}
\end{equation}
denotes the completion time of column $S$.

Equation~\eqref{eq:rmp-partition} ensures that each vehicle is covered exactly once by the selected columns, i.e., every vehicle is assigned one feasible candidate interval.
Equation~\eqref{eq:rmp-makespan} links the makespan variable $\tau$ with the completion time of each selected column. Since $T_S$ is defined in Equation~\eqref{eq:rmp-ts}, this constraint guarantees that $\tau$ is no smaller than the completion time induced by any chosen column.
Equation~\eqref{eq:rmp-conflict} enforces charger-wise conflict exclusion: for any conflict edge $(u,v)\in\mathcal{E}^A$, two conflicting vertices cannot be selected simultaneously on the same charger $c$.
Equation~\eqref{eq:rmp-nonneg} imposes nonnegativity on the decision variables.

\bigskip
\noindent\textbf{Pricing Subproblem}

The pricing subproblem is solved using dual information from the RMP.  
Introduce binary variables
\begin{equation}
x_{v,c}=
\begin{cases}
1, & \text{if vertex $v$ is selected on charger $c$,}\\
0, & \text{otherwise,}
\end{cases}
\qquad \forall v\in\mathcal{I},\ c\in\mathcal{C}.
\end{equation}

The pricing problem is formulated as
\begin{equation}
\max_{x}\ \sum_{v\in\mathcal{I}}\sum_{c\in\mathcal{C}} w_v x_{v,c}
\label{eq:pricing-obj-max}
\end{equation}
subject to
\begin{equation}
\sum_{v\in P_n}\sum_{c\in\mathcal{C}} x_{v,c}=1,
\quad \forall n\in N,
\label{eq:pricing-partition}
\end{equation}
\begin{equation}
x_{u,c}+x_{v,c}\le 1,
\quad \forall (u,v)\in\mathcal{E}^A,\ \forall c\in\mathcal{C},
\label{eq:pricing-conflict}
\end{equation}
\begin{equation}
x_{v,c}\in\{0,1\},
\quad \forall v\in\mathcal{I},\forall c\in\mathcal{C}.
\label{eq:pricing-binary}
\end{equation}

Equation~\eqref{eq:pricing-partition} enforces that each vehicle $n$ selects exactly one candidate interval across all chargers.
Equation~\eqref{eq:pricing-conflict} enforces charger-wise conflict exclusion.
Equation~\eqref{eq:pricing-binary} defines integrality.

The selected vertices induce a candidate independent set
\[
S=\{v\in\mathcal{I}\mid x_{v,c}=1\ \text{for some }c\in\mathcal{C}\}.
\]
A new column is added to the RMP when the pricing objective indicates a negative reduced-cost column.

\noindent\textbf{QUBO Formulation for the Pricing Subproblem}

The pricing subproblem can also be expressed as a QUBO model:
\begin{equation}
\mathcal{H}_0 = -\sum_{v \in \mathcal{I}} \sum_{c \in \mathcal{C}} w_v \, x_{v,c},
\end{equation}
\begin{equation}
\mathcal{H}_1 = \sum_{n \in N} \rho_1 \left( \sum_{v \in P_n} \sum_{c \in \mathcal{C}} x_{v,c} - 1 \right)^2,
\end{equation}
\begin{equation}
\mathcal{H}_2 = \sum_{c \in \mathcal{C}} \sum_{(u,v) \in \mathcal{E}^A} \rho_2 \, x_{u,c} x_{v,c},
\end{equation}
\begin{equation}
\mathcal{H} = \mathcal{H}_0 + \mathcal{H}_1 + \mathcal{H}_2.
\end{equation}

To improve the quality of heuristic pricing, we optionally tune the QUBO and solver parameters before generating columns. In our implementation, this automatic tuning is performed with Optuna. The search space includes the penalty coefficients $(\rho_1,\rho_2)$ and the main QAIA settings, such as the time step, scaling factor, iteration budget, batch size, and restart count. We first build the QUBO with the sampled $(\rho_1,\rho_2)$, then run the QAIA heuristic to obtain a binary assignment, apply greedy repair when needed, and evaluate the resulting column quality. The best configuration is then used to generate candidate columns, and only columns with negative reduced cost are added to the RMP.

Solving this QUBO yields a binary assignment $x_{v,c}\in\{0,1\}$.
A feasible independent set is then obtained by aggregating selected vertices over chargers, with greedy repair applied when necessary.
\bigskip

\noindent\textbf{Node Selection Strategy}

Before branching, a best-bound node selection strategy is adopted. At each iteration, the pending node with the smallest LP relaxation objective value is selected for expansion. If multiple nodes share the same value, their processing order is determined by the priority queue.
\subsection{Branching Scheme}

To explore the solution space efficiently, we use a two-stage branching scheme that is consistent with both the conflict graph and the vehicle-partition structure. The two rules are subset branching and pairwise branching \cite{zykov1952some}.

\noindent\textbf{Subset Branching}

At a given node, if a partition $P_n$ contains more than one selected vertex in the current RMP solution,
\begin{equation}
\left|\left\{v \in P_n : \sum_{S \in \mathcal{S}':\, v \in S}\sum_{c\in\mathcal{C}} \zeta_{S,c}^* > 0 \right\}\right| > 1,
\end{equation}
we select the partition $P_{n^\star}$ with the largest number of selected vertices. Within $P_{n^\star}$, the vertex
\begin{equation}
v^\star = \arg\max_{v \in P_{n^\star}} \sum_{S \in \mathcal{S}':\, v \in S}\sum_{c\in\mathcal{C}} \zeta_{S,c}^*
\end{equation}
is selected for branching.

Two complementary child nodes are generated:

\begin{itemize}
    \item \textbf{Select branch:} vertex $v^\star$ is fixed as the chosen interval for vehicle $n^\star$, and all other vertices in $P_{n^\star}\setminus\{v^\star\}$ are excluded from further consideration.
    
    \item \textbf{Discard branch:} vertex $v^\star$ is forbidden and removed from further consideration.
\end{itemize}

This rule drives each partition toward a single chosen interval while preserving the overall pricing structure.

\noindent\textbf{Pairwise Branching}

If the RMP solution is still fractional after subset branching, we branch on a pair of vertices from different partitions. Specifically, for a pair $(u,v)$ with $u$ and $v$ belonging to different partitions, define
\begin{equation}
\gamma(u,v) = \sum_{S \in \mathcal{S}':\, u,v \in S}\sum_{c\in\mathcal{C}} \zeta_{S,c}^*.
\end{equation}
We select the pair $(u^\star,v^\star)$ such that $\gamma(u^\star,v^\star)$ is non-integer and maximized over all cross-partition pairs.

Two child nodes are created:

\begin{itemize}
    \item \textbf{Different color branch:} enforce that $u^\star$ and $v^\star$ cannot be selected together. This is implemented by adding an edge $(u^\star,v^\star)$ to the conflict graph.
    
    \item \textbf{Same color branch:} enforce that $u^\star$ and $v^\star$ must be selected together. A new super vertex $z$ is created to replace $u^\star$ and $v^\star$, with edges
    \begin{equation}
    (z,w) \in \mathcal{E}^A \quad \text{if } (u^\star,w) \in \mathcal{E}^A \text{ or } (v^\star,w) \in \mathcal{E}^A.
    \end{equation}
    The two vertices are merged in the auxiliary graph through a new combined vertex representation.
\end{itemize}

\noindent\textbf{Fathoming Criteria and Completeness}

A node is fathomed if it is infeasible or if its lower bound exceeds the current global upper bound. By applying the two rules sequentially, the algorithm eliminates fractional RMP solutions and enforces one selected interval per vehicle. Importantly, the pricing subproblem keeps the same structure throughout the tree, which preserves both consistency and completeness of the branching process.

\section{Computational results}\label{sec:Experimental}
\subsection{Instances}\label{sec:Instance}
All experiments were conducted on a Windows~10 PC with a 12th-generation Intel\textsuperscript{\textregistered} Core\textsuperscript{\texttrademark} i7-12700H processor, using Python~3.10.18 (64-bit). A set of benchmark instances for the PCP was generated based on the EV charging scheduling scenario. Let $T$ denote the scheduling horizon, $d$ the fixed charging duration, $|N|$ the number of vehicles, and $C$ the number of charging piles. For each vehicle $i \in N$, $K$ candidate charging start times $s_{i,k}$ were sampled uniformly at random from $[0,\,T-d]$, yielding intervals $[s_{i,k},\,s_{i,k}+d]$. By fixing $|\mathcal I|$ (the total number of intervals) and $K$, the number of vehicles is $|N| = |\mathcal I|/K$. Each interval corresponds to a vertex in the PCP graph, and vertices of the same vehicle form a partition. Conflict edges connect vertices either within the same partition or between overlapping intervals of different vehicles, respecting the charging pile capacity $C$.

All instances used $T=24$, $d=3$, and $C \in \{5,10\}$. They were stored in text format with metadata recording time windows and partition indices. Files follow the naming convention \texttt{V\{V\}\_vpc\{K\}\_s\{S\}} (e.g., \texttt{V120\_vpc6\_s1}), where $V$ indicates the total number of candidate intervals or PCP vertices (i.e., $|\mathcal I|$), \texttt{vpc} denotes the number of candidate intervals per vehicle ($K$), and $S$ is the random seed. Under this convention, the number of vehicles is $|N|=|\mathcal I|/K$. The maximum computational time per instance was 3600 seconds; if this limit was reached, the solver returned the incumbent best feasible solution.

\subsection{Numerical results}
\label{subsec:numerical_results}

In this subsection, we numerically evaluate the proposed branch-and-price framework on the EV charging PCP instances described in Section~\ref{sec:Instance}. We compare three variants of the algorithm:
\begin{enumerate}
        \item \textbf{Gurobi baseline}: both the RMP and the pricing subproblem are solved exactly by the classical MIP solver Gurobi.
    \item \textbf{bSB}: the RMP is still solved by Gurobi, while the pricing subproblem is solved heuristically by the ballistic simulated branching algorithm \texttt{mindquantum.algorithm.qaia.bSB}.
    \item \textbf{SimCIM}: the RMP is solved by Gurobi and the pricing subproblem is solved heuristically by the simulated coherent Ising machine algorithm \texttt{mindquantum.algorithm.qaia.SimCIM}.
\end{enumerate}
For brevity, we refer to the two QAIA variants simply as bSB and SimCIM in what follows. All numerical simulations presented in this work were implemented using the MindSpore Quantum framework \cite{xu2024mindspore}, which provides a powerful environment for quantum computation simulation. The source code and data required to reproduce the numerical results reported in this paper are publicly available at \url{https://github.com/Liang023/BP_PCP_Cn}.

\begin{figure}[H]
    \centering

    \begin{minipage}{0.48\linewidth}
        \centering
        \includegraphics[width=\linewidth]{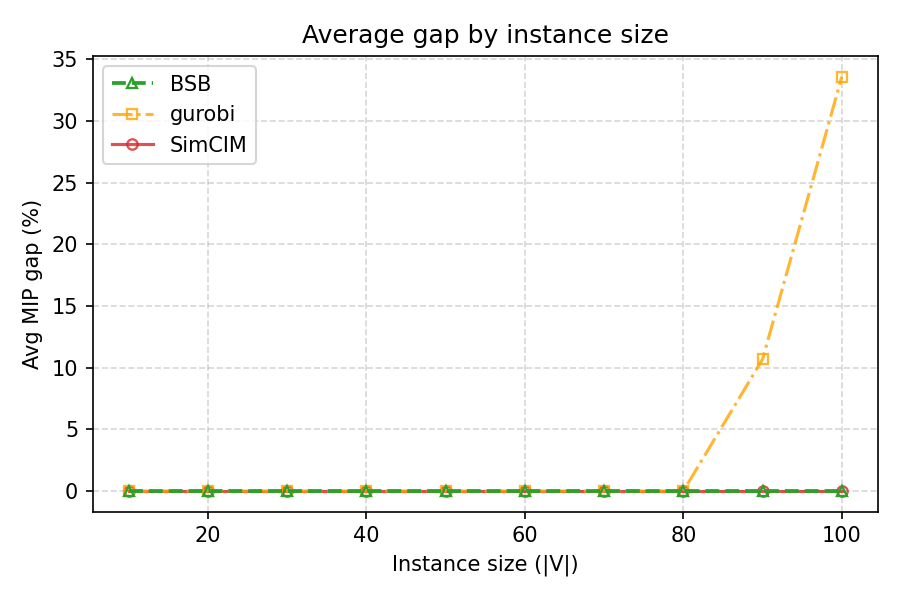}
        {\scriptsize (a)}
    \end{minipage}
    \hfill
    \begin{minipage}{0.48\linewidth}
        \centering
        \includegraphics[width=\linewidth]{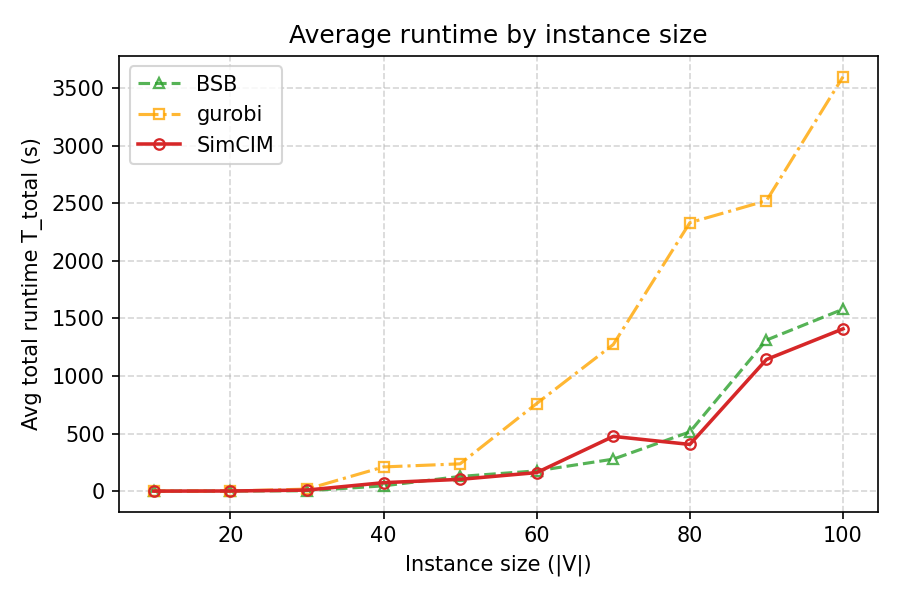}
        {\scriptsize (b)}
    \end{minipage}

    \caption{Scalability comparison of branch-and-price variants. 
    Subfigure (a) MIP gap versus instance size, where QAIA methods bSB and SimCIM consistently achieve zero optimality gap across all instances, while the Gurobi baseline exhibits rapidly increasing gaps for large-scale cases. 
    Subfigure (b) Total runtime versus instance size, where all methods perform similarly on small instances, but bSB and SimCIM scale significantly better than the baseline as problem size increases.}

    \label{fig:f5_f6}
\end{figure}

For each instance and each method, Tables~\ref{tab:gurobi}--\ref{tab:SimCIM} report the following performance indicators. Columns $|\mathcal I|$, $|\mathcal E|$, and $|N|$ denote the number of candidate intervals or vertices, the number of edges, and the number of vehicles, respectively. Here, $|\mathcal I|$ matches the first numeric field in the instance name. Column \texttt{Obj.} gives the best objective value at termination, and \texttt{Gap (\%)} is the relative optimality gap reported by the solver; a value of zero indicates that global optimality has been proved. The column $T_{\mathrm{total}}$ reports the wall clock time in seconds spent inside the branch-and-price procedure, from the construction of the initial RMP until the last node is processed; it includes all calls to the MIP solver and the control logic of column generation, but excludes instance reading and final printing.

We further decompose the solver time into two components: $T_{\mathrm{RMP}}$ is the cumulative time spent solving all RMPs, and $T_{\mathrm{pricing}}$ is the cumulative time spent solving all pricing subproblems. Finally, $N_p$ denotes the number of pricing calls, $N_c$ the number of generated columns, and $N_n$ the number of processed branch-and-price nodes. Tables~\ref{tab:gurobi}, \ref{tab:bSB}, and~\ref{tab:SimCIM} report these quantities for the Gurobi baseline, bSB, and SimCIM, respectively.

\begin{table}[!htbp]
\centering
\caption{Computational results of the instances solved by Gurobi}
\label{tab:gurobi}
\renewcommand{\arraystretch}{1.1}
\resizebox{\linewidth}{!}{
\begin{tabular}{lrrrrrrrrrrr}
\toprule
Instance & $|\mathcal I|$ & $|\mathcal E|$ & $|N|$ & Obj. & Gap (\%) &
$T_{\mathrm{total}}$ &
$T_{\mathrm{RMP}}$ &
$T_{\mathrm{pricing}}$ &
$N_p$ & $N_c$ & $N_n$ \\
\midrule
v10c5k2s1   & 10 & 11  & 5  & 18 & 0.00 & 0.59 & 0.04 & 0.52 & 39   & 1\,224    & 10 \\
v10c5k2s2   & 10 & 15  & 5  & 21 & 0.00 & 0.49 & 0.04 & 0.43 & 34   & 2\,310    & 8  \\
v10c5k2s3   & 10 & 22  & 5  & 21 & 0.00 & 0.76 & 0.06 & 0.67 & 63   & 3\,868    & 17 \\

v20c5k2s1   & 20 & 39  & 10 & 18 & 0.00 & 0.71 & 0.23 & 0.41 & 36   & 4\,587    & 12 \\
v20c5k2s2   & 20 & 46  & 10 & 21 & 0.00 & 0.69 & 0.19 & 0.46 & 29   & 8\,263    & 9  \\
v20c5k2s3   & 20 & 57  & 10 & 21 & 0.00 & 0.75 & 0.16 & 0.53 & 24   & 11\,321   & 8  \\

v30c5k3s1   & 30 & 118 & 10 & 17 & 0.00 & 11.89 & 3.30 & 7.80 & 252  & 50\,607   & 77 \\
v30c5k3s2   & 30 & 108 & 10 & 17 & 0.00 & 29.02 & 8.91 & 17.41 & 1\,295 & 201\,841  & 461 \\
v30c5k3s3   & 30 & 123 & 10 & 18 & 0.00 & 11.92 & 3.77 & 7.15 & 325  & 262\,858  & 104 \\

v40c5k4s1   & 40 & 205 & 10 & 15 & 0.00 & 346.50 & 139.39 & 167.50 & 3\,229 & 597\,481  & 1\,055 \\
v40c5k4s2   & 40 & 219 & 10 & 15 & 0.00 & 169.20 & 64.38 & 88.38 & 1\,626 & 922\,675  & 495 \\
v40c5k4s3   & 40 & 197 & 10 & 15 & 0.00 & 114.99 & 53.62 & 47.47 & 1\,037 & 1\,146\,298 & 321 \\

v50c10k5s1  & 50 & 320 & 10 & 10 & 0.00 & 136.28 & 53.89 & 74.26 & 744  & 305\,519  & 230 \\
v50c10k5s2  & 50 & 342 & 10 & 14 & 0.00 & 366.65 & 153.63 & 183.80 & 4\,158 & 1\,354\,840 & 1\,306 \\
v50c10k5s3  & 50 & 338 & 10 & 12 & 0.00 & 208.38 & 80.11 & 116.46 & 1\,295 & 1\,819\,533 & 382 \\

v60c10k6s1  & 60 & 490 & 10 & 10 & 0.00 & 259.33 & 105.56 & 138.27 & 1\,444 & 486\,050  & 467 \\
v60c10k6s2  & 60 & 509 & 10 & 14 & 0.00 & 1\,047.50 & 461.93 & 505.34 & 10\,084 & 3\,060\,585 & 3\,305 \\
v60c10k6s3  & 60 & 512 & 10 & 10 & 0.00 & 976.85 & 377.42 & 546.14 & 2\,953 & 1\,156\,494 & 856 \\

v70c10k7s1  & 70 & 696 & 10 & 10 & 0.00 & 2\,674.72 & 1\,248.74 & 1\,236.48 & 17\,403 & 5\,463\,412 & 5\,557 \\
v70c10k7s2  & 70 & 710 & 10 & 10 & 0.00 & 406.67 & 183.62 & 203.44 & 1\,358 & 6\,095\,377 & 396 \\
v70c10k7s3  & 70 & 679 & 10 & 12 & 0.00 & 740.31 & 352.93 & 352.04 & 2\,021 & 7\,337\,643 & 542 \\

v80c10k8s1  & 80 & 918 & 10 & 6  & 0.00 & 2\,367.23 & 1\,318.34 & 731.75 & 2\,829 & 1\,875\,064 & 861 \\
v80c10k8s2  & 80 & 904 & 10 & 11 & 0.00 & 1\,472.33 & 768.11 & 598.06 & 6\,492 & 4\,304\,470 & 2\,161 \\
v80c10k8s3  & 80 & 891 & 10 & 12 & 0.00 & 3\,158.11 & 1\,572.39 & 1\,410.68 & 9\,101 & 9\,114\,941 & 2\,719 \\

v90c10k9s1  & 90 & 1\,164 & 10 & 10 & 32.08 & 3\,600.02 & 1\,827.34 & 1\,623.28 & 5\,513 & 4\,062\,436 & 1\,664 \\
v90c10k9s2  & 90 & 1\,137 & 10 & 11 & 0.00  & 820.15 & 312.43 & 490.66 & 249 & 4\,509\,936 & 62 \\
v90c10k9s3  & 90 & 1\,144 & 10 & 6  & 0.00  & 3\,149.07 & 1\,603.51 & 1\,407.98 & 4\,801 & 3\,477\,710 & 1\,350 \\

v100c10k10s1 & 100 & 1\,437 & 10 & 14 & 35.71 & 3\,600.47 & 1\,870.14 & 1\,563.06 & 4\,849 & 3\,619\,512 & 1\,469 \\
v100c10k10s2 & 100 & 1\,372 & 10 & 12 & 25.00 & 3\,600.07 & 2\,209.08 & 1\,220.32 & 4\,185 & 7\,291\,623 & 1\,280 \\
v100c10k10s3 & 100 & 1\,398 & 10 & 15 & 40.00 & 3\,600.31 & 1\,551.81 & 1\,955.20 & 368 & 8\,098\,615 & 97 \\
\bottomrule
\end{tabular}}
\end{table}

\begin{table}[!htbp]
\centering
\caption{Computational results of instances solved by the bSB method}
\label{tab:bSB}
\renewcommand{\arraystretch}{1.1}
\resizebox{\linewidth}{!}{
\begin{tabular}{lrrrrrrrrrrr}
\toprule
Instance 
& $|\mathcal I|$ 
& $|\mathcal E|$ 
& $|N|$ 
& Obj. 
& Gap (\%) 
& $T_{\text{total}}$ 
& $T_{\text{RMP}}$ 
& $T_{\text{pricing}}$ 
& $N_p$ 
& $N_c$ 
& $N_n$ \\
\midrule
v10c5k2s1   & 10 & 11  & 5  & 18 & 0.00 & 0.13 & 0.03 & 0.08 & 35 & 891 & 8 \\
v10c5k2s2   & 10 & 15  & 5  & 21 & 0.00 & 0.12 & 0.03 & 0.08 & 34 & 1\,608 & 10 \\
v10c5k2s3   & 10 & 22  & 5  & 21 & 0.00 & 0.11 & 0.03 & 0.07 & 33 & 2\,211 & 8 \\
v20c5k2s1   & 20 & 39  & 10 & 18 & 0.00 & 2.07 & 1.04 & 0.80 & 60 & 6\,154 & 18 \\
v20c5k2s2   & 20 & 46  & 10 & 21 & 0.00 & 0.54 & 0.28 & 0.18 & 34 & 9\,808 & 12 \\
v20c5k2s3   & 20 & 57  & 10 & 21 & 0.00 & 0.32 & 0.17 & 0.11 & 29 & 13\,123 & 10 \\
v30c5k3s1   & 30 & 118 & 10 & 17 & 0.00 & 3.47 & 2.26 & 0.75 & 197 & 31\,870 & 72 \\
v30c5k3s2   & 30 & 108 & 10 & 17 & 0.00 & 4.07 & 2.55 & 0.98 & 266 & 69\,548 & 105 \\
v30c5k3s3   & 30 & 123 & 10 & 18 & 0.00 & 2.25 & 1.52 & 0.46 & 98 & 90\,877 & 37 \\
v40c5k4s1   & 40 & 205 & 10 & 15 & 0.00 & 10.34 & 7.52 & 1.55 & 79 & 33\,518 & 30 \\
v40c5k4s2   & 40 & 219 & 10 & 15 & 0.00 & 24.34 & 16.66 & 4.15 & 215 & 96\,163 & 82 \\
v40c5k4s3   & 40 & 197 & 10 & 15 & 0.00 & 102.66 & 62.39 & 25.15 & 1\,895 & 402\,235 & 728 \\
v50c10k5s1  & 50 & 320 & 10 & 10 & 0.00 & 54.79 & 40.01 & 9.25 & 909 & 233\,195 & 337 \\
v50c10k5s2  & 50 & 342 & 10 & 14 & 0.00 & 161.77 & 116.58 & 27.57 & 2\,401 & 814\,713 & 934 \\
v50c10k5s3  & 50 & 338 & 10 & 12 & 0.00 & 166.67 & 126.93 & 21.10 & 393 & 999\,817 & 151 \\
v60c10k6s1  & 60 & 490 & 10 & 10 & 0.00 & 97.52 & 72.02 & 15.44 & 1\,870 & 320\,294 & 759 \\
v60c10k6s2  & 60 & 509 & 10 & 14 & 0.00 & 334.88 & 249.28 & 50.21 & 4\,704 & 1\,605\,097 & 1\,868 \\
v60c10k6s3  & 60 & 512 & 10 & 10 & 0.00 & 95.43 & 77.96 & 9.14 & 447 & 1\,834\,619 & 177 \\
v70c10k7s1  & 70 & 696 & 10 & 10 & 0.00 & 428.59 & 336.82 & 53.66 & 4\,467 & 1\,163\,328 & 1\,758 \\
v70c10k7s2  & 70 & 710 & 10 & 10 & 0.00 & 159.92 & 136.59 & 12.06 & 499 & 1\,478\,417 & 184 \\
v70c10k7s3  & 70 & 679 & 10 & 12 & 0.00 & 245.35 & 211.69 & 18.39 & 687 & 1\,962\,177 & 273 \\
v80c10k8s1  & 80 & 918 & 10 & 6  & 0.00 & 283.31 & 250.68 & 16.46 & 518 & 438\,516 & 195 \\
v80c10k8s2  & 80 & 904 & 10 & 11 & 0.00 & 406.65 & 342.71 & 34.52 & 1\,860 & 1\,298\,232 & 743 \\
v80c10k8s3  & 80 & 891 & 10 & 12 & 0.00 & 854.50 & 727.35 & 69.19 & 2\,754 & 3\,004\,040 & 1\,072 \\
v90c10k9s1  & 90 & 1\,164 & 10 & 7  & 0.00 & 1\,241.05 & 1\,115.71 & 66.23 & 1\,748 & 1\,573\,060 & 688 \\
v90c10k9s2  & 90 & 1\,137 & 10 & 11 & 0.00 & 853.65 & 762.13 & 42.38 & 997 & 2\,623\,423 & 392 \\
v90c10k9s3  & 90 & 1\,144 & 10 & 6  & 0.00 & 1\,842.04 & 1\,588.64 & 130.23 & 4\,625 & 2\,969\,128 & 1\,736 \\
v100c10k10s1 & 100 & 1\,437 & 10 & 10 & 0.00 & 965.76 & 844.34 & 62.93 & 4\,616 & 1\,304\,499 & 1\,844 \\
v100c10k10s2 & 100 & 1\,372 & 10 & 10 & 0.00 & 3\,217.70 & 2\,707.04 & 282.36 & 17\,780 & 7\,342\,087 & 7\,027 \\
v100c10k10s3 & 100 & 1\,398 & 10 & 10 & 0.00 & 558.98 & 517.09 & 18.68 & 383 & 7\,832\,319 & 152 \\
\bottomrule
\end{tabular}}
\end{table}

\begin{table}[!htbp]
\centering
\caption{Computational results of instances solved by the SimCIM method}
\label{tab:SimCIM}
\renewcommand{\arraystretch}{1.1}
\resizebox{\linewidth}{!}{
\begin{tabular}{lrrrrrrrrrrr}
\toprule
Instance 
& $|\mathcal I|$ 
& $|\mathcal E|$ 
& $|N|$ 
& Obj. 
& Gap (\%) 
& $T_{\text{total}}$ 
& $T_{\text{RMP}}$ 
& $T_{\text{pricing}}$ 
& $N_p$ 
& $N_c$ 
& $N_n$ \\
\midrule
v10c5k2s1   & 10 & 11   & 5  & 18 & 0.00 & 0.14 & 0.04 & 0.09 & 36   & 950      & 7 \\
v10c5k2s2   & 10 & 15   & 5  & 21 & 0.00 & 0.18 & 0.05 & 0.13 & 52   & 2\,011   & 10 \\
v10c5k2s3   & 10 & 22   & 5  & 21 & 0.00 & 0.15 & 0.04 & 0.10 & 40   & 2\,765   & 8 \\
v20c5k2s1   & 20 & 39   & 10 & 18 & 0.00 & 0.95 & 0.46 & 0.42 & 111  & 11\,406  & 23 \\
v20c5k2s2   & 20 & 46   & 10 & 21 & 0.00 & 1.60 & 0.73 & 0.72 & 182  & 27\,629  & 47 \\
v20c5k2s3   & 20 & 57   & 10 & 21 & 0.00 & 0.80 & 0.38 & 0.37 & 85   & 36\,912  & 13 \\
v30c5k3s1   & 30 & 118  & 10 & 17 & 0.00 & 9.42 & 5.03 & 3.84 & 672  & 140\,893 & 60 \\
v30c5k3s2   & 30 & 108  & 10 & 17 & 0.00 & 4.20 & 2.26 & 1.70 & 280  & 203\,937 & 29 \\
v30c5k3s3   & 30 & 123  & 10 & 18 & 0.00 & 13.81 & 7.18 & 6.06 & 1\,001 & 419\,810 & 61 \\
v40c5k4s1   & 40 & 205  & 10 & 15 & 0.00 & 69.15 & 38.26 & 26.92 & 4\,859 & 1\,054\,133 & 567 \\
v40c5k4s2   & 40 & 219  & 10 & 15 & 0.00 & 69.41 & 39.12 & 26.03 & 4\,518 & 2\,085\,768 & 497 \\
v40c5k4s3   & 40 & 197  & 10 & 15 & 0.00 & 81.63 & 44.23 & 32.15 & 6\,121 & 3\,309\,729 & 887 \\
v50c10k5s1  & 50 & 320  & 10 & 10 & 0.00 & 39.62 & 27.81 & 8.22 & 503  & 179\,936 & 125 \\
v50c10k5s2  & 50 & 342  & 10 & 14 & 0.00 & 183.99 & 121.77 & 47.44 & 2\,998 & 1\,140\,887 & 621 \\
v50c10k5s3  & 50 & 338  & 10 & 12 & 0.00 & 83.81 & 58.07 & 19.03 & 1\,219 & 1\,536\,352 & 295 \\
v60c10k6s1  & 60 & 490  & 10 & 10 & 0.00 & 68.95 & 47.99 & 14.81 & 1\,596 & 241\,241 & 505 \\
v60c10k6s2  & 60 & 509  & 10 & 14 & 0.00 & 309.76 & 218.84 & 65.62 & 3\,885 & 1\,523\,498 & 1\,121 \\
v60c10k6s3  & 60 & 512  & 10 & 10 & 0.00 & 104.12 & 82.42 & 13.21 & 596   & 1\,796\,870 & 200 \\
v70c10k7s1  & 70 & 696  & 10 & 10 & 0.00 & 767.03 & 582.54 & 123.22 & 7\,099 & 2\,424\,686 & 2\,234 \\
v70c10k7s2  & 70 & 710  & 10 & 10 & 0.00 & 411.35 & 328.28 & 52.47 & 2\,400 & 3\,466\,319 & 797 \\
v70c10k7s3  & 70 & 679  & 10 & 12 & 0.00 & 248.01 & 197.73 & 37.25 & 1\,039 & 4\,314\,303 & 194 \\
v80c10k8s1  & 80 & 918  & 10 & 6  & 0.00 & 441.98 & 374.12 & 39.50 & 1\,794 & 787\,934 & 682 \\
v80c10k8s2  & 80 & 904  & 10 & 11 & 0.00 & 444.22 & 363.58 & 49.50 & 2\,569 & 1\,822\,835 & 883 \\
v80c10k8s3  & 80 & 891  & 10 & 12 & 0.00 & 331.01 & 287.02 & 26.55 & 698   & 2\,454\,506 & 209 \\
v90c10k9s1  & 90 & 1\,164 & 10 & 7  & 0.00 & 742.89 & 650.83 & 50.88 & 1\,387 & 1\,093\,815 & 493 \\
v90c10k9s2  & 90 & 1\,137 & 10 & 11 & 0.00 & 853.12 & 756.72 & 47.85 & 1\,002 & 2\,149\,515 & 392 \\
v90c10k9s3  & 90 & 1\,144 & 10 & 6  & 0.00 & 1\,834.94 & 1\,567.18 & 147.81 & 4\,646 & 5\,130\,084 & 1\,711 \\
v100c10k10s1 & 100 & 1\,437 & 10 & 10 & 0.00 & 968.25 & 835.83 & 74.32 & 5\,049 & 1\,344\,067 & 1\,893 \\
v100c10k10s2 & 100 & 1\,372 & 10 & 10 & 0.00 & 2\,726.95 & 2\,380.15 & 183.05 & 4\,633 & 5\,508\,556 & 1\,759 \\
v100c10k10s3 & 100 & 1\,398 & 10 & 10 & 0.00 & 534.31 & 491.69 & 20.21 & 383   & 5\,998\,788 & 152 \\
\bottomrule
\end{tabular}}
\end{table}

On \emph{small and medium} instances ($|\mathcal I| \le 40$ with $K \in \{2,3,4\}$), all three methods can find and prove the global optimum, i.e., \texttt{gap}~=~0 for all instances. The total runtime $T_{\mathrm{total}}$ is in the sub-second to tens-of-seconds range for all methods. In this regime, replacing the classical Gurobi pricing with bSB or SimCIM does not deteriorate solution quality and leads to comparable running times. Small variations in $T_{\mathrm{total}}$ mainly stem from differences in the number of pricing calls and the efficiency of the respective pricing solvers, but overall, all three methods behave similarly on these easier instances.

The benefit of QAIA pricing becomes pronounced on \emph{large and hard} instances ($|\mathcal I| \ge 80$ with $K \in \{8,9,10\}$). For several of the largest instances, the pure Gurobi baseline hits the time limit of $3600$~s and terminates with a nonzero optimality gap up to about $35$--$40\%$ on some $|\mathcal I|=100$ instances, as shown in Table~\ref{tab:gurobi}. In contrast, both bSB and SimCIM are able to close the gap and prove optimality within the same time limit, often with a significantly smaller total runtime. For example, for instance \texttt{v100c10k10s1}, the baseline method stops at $T_{\mathrm{total}} \approx 3600$~s with a large remaining gap, whereas bSB and SimCIM reach the optimal solution with \texttt{gap}~=~0 in roughly $900$--$1000$~s. Similar patterns can be observed in other $|\mathcal I|=80,90,100$ instances: the two QAIA methods consistently match the best objective values of the baseline and, in many of the hardest cases, convert time limit with gap runs of Gurobi into fully optimal solutions.

The overall trends are further illustrated in Figure~\ref{fig:f5_f6}, panels a and b, which plot the MIP gap and total runtime as a function of the instance size $|\mathcal I|$ for the three methods. Panel a shows that bSB and SimCIM achieve \texttt{gap}~=~0 on all tested instances, while the Gurobi baseline exhibits large positive gaps only on the largest instances. Panel b indicates that all methods have similar runtimes on small instances, but as $|\mathcal I|$ grows, the total runtime of the baseline method increases much faster than that of the QAIA methods. In particular, for the largest instances, the bSB and SimCIM curves stay well below the $3600$~s time limit, whereas the baseline curve often saturates at the time limit.

These observations suggest that QAIA pricing improves the effectiveness of the column generation and branching process. Although the detailed internal statistics are not all shown in the figures, Tables~\ref{tab:gurobi}--\ref{tab:SimCIM} and the logs indicate that bSB and SimCIM typically generate a large number of high quality columns within a moderate number of pricing calls, enabling the RMP to capture the structure of the problem more quickly and allowing the branch-and-price tree to be explored with fewer or comparable processed nodes while still certifying optimality. Comparing the two QAIA variants, bSB generally achieves slightly smaller or comparable $T_{\mathrm{total}}$ on many of the larger instances, while SimCIM provides similar solution quality and often comparable performance, acting as a complementary quantum-inspired heuristic. Overall, embedding MindQuantum’s QAIA solvers, realized here as bSB and SimCIM, into the pricing subproblem of a classical branch-and-price framework yields a robust hybrid algorithm: on small and medium instances, it matches the performance of the pure Gurobi baseline, and on large and difficult instances, it significantly improves scalability and robustness by delivering high-quality, often provably optimal, solutions within much shorter runtimes than the classical baseline.

\section{Conclusions}\label{sec:Conclusions}

In this manuscript, we studied an intraday electric vehicle charging scheduling problem and formulated it as a PCP. Vehicles are modeled as partitions, candidate charging intervals as vertices, and conflict edges represent temporal overlap, resource incompatibility, and same-partition exclusivity in a conflict graph. This representation makes the one interval per vehicle requirement and charger capacity constraints explicit and is well-suited to instances with many alternative intervals.

Building on this formulation, we propose a hybrid branch-and-price algorithm that combines classical exact optimization with quantum-inspired heuristics. The RMP is solved by Gurobi, while the pricing subproblem is cast as a maximum independent set problem and further reformulated as a QUBO model to be tackled by QAIA. Two QAIA pricing schemes, using bSB and SimCIM, were implemented within the MindQuantum framework on classical hardware and compared with a pure Gurobi baseline branch-and-price method.

Numerical experiments on synthetic benchmark instances show that all three methods obtain optimal solutions with comparable runtimes on small and medium problems. For large and difficult instances, however, the QAIA enhanced variants achieve markedly better scalability: they often close optimality gaps and prove optimality within the time limit in cases where the baseline hits the limit with sizable gaps. Overall, the results suggest that embedding QAIA in classical decomposition algorithms is an effective way to improve the practical solvability of large-scale EV charging scheduling and related PCP models. Future work will consider richer network constraints, data-driven instance generation, and the integration of other quantum-inspired or variational algorithms within the same framework.

\section*{Acknowledgments}
This work was sponsored by CPS-Yangtze Delta Region Industrial Innovation Center of Quantum and Information Technology -- MindSpore Quantum Open Fund.
\bibliographystyle{unsrt}  
\bibliography{references} 

@article{liu2026digital,
  title={A digital twin framework for intelligent electric vehicle charging optimization in smart manufacturing systems},
  author={Liu, Chunting and Liu, Ruyu and Liu, Xiufeng},
  journal={Applied Energy},
  volume={406},
  pages={127281},
  year={2026},
  publisher={Elsevier}
}

@article{fang2026game,
  title={Game-based scheduling of mobile charging robots for electric vehicle charging: A relay-like scheme},
  author={Fang, Qiuyang and Zhang, Chunyan and Wang, Chen and Xie, Guangming and Zhang, Jianlei},
  journal={Applied Energy},
  volume={402},
  pages={126956},
  year={2026},
  publisher={Elsevier}
}

@article{huang2026health,
  title={Health-aware proportional-fair scheduling of fast electric vehicle charging stations under distribution-level power envelopes},
  author={Huang, Min and Wang, Haoyu},
  journal={Electric Power Systems Research},
  volume={257},
  pages={112944},
  year={2026},
  publisher={Elsevier}
}

@article{khiar2025multi,
  title={Multi-objective electric vehicle charging scheduling under stochastic duration uncertainty},
  author={Khiar, Aimen and el Amine Brahmia, Mohamed and Oulamara, Ammar and Idoumghar, Lhassane},
  journal={Omega},
  pages={103506},
  year={2025},
  publisher={Elsevier}
}

@article{gauchotte2026study,
  title={Study of Electric Vehicle Charging Scheduling with Renewable Energy: Offline and Stochastic Online Optimization},
  author={Gauchotte, R and Oulamara, A and Ghogho, M and Oudani, M},
  journal={European Journal of Operational Research},
  year={2026},
  publisher={Elsevier}
}

@article{tiunov2019annealing,
  title={Annealing by simulating the coherent Ising machine},
  author={Tiunov, Egor S and Ulanov, Alexander E and Lvovsky, AI},
  journal={Optics express},
  volume={27},
  number={7},
  pages={10288--10295},
  year={2019},
  publisher={Optical Society of America}
}

@misc{xu2024mindspore,
title={MindSpore
Quantum:
A
User-Friendly,
High-Performance, and AI-Compatible Quantum Computing
Framework},
author={Xusheng Xu and Jiangyu Cui and Zidong Cui and
Runhong He and Qingyu Li and Xiaowei Li and Yanling Lin and
Jiale Liu and Wuxin Liu and Jiale Lu and others},
year={2024},
eprint={2406.17248},
archivePrefix={arXiv},
primaryClass={quant-ph},
url={https://arxiv.org/abs/2406.17248},
}

@book{zykov1952some,
  title={On some properties of linear complexes},
  author={Zykov, Aleksandr Aleksandrovich},
  number={79},
  year={1952},
  publisher={American Mathematical Society}
}

@article{zeng2024performance,
  title={Performance of quantum annealing inspired algorithms for combinatorial optimization problems},
  author={Zeng, Qing-Guo and Cui, Xiao-Peng and Liu, Bowen and Wang, Yao and Mosharev, Pavel and Yung, Man-Hong},
  journal={Communications Physics},
  volume={7},
  number={1},
  pages={249},
  year={2024},
  publisher={Nature Publishing Group UK London}
}

@article{kanao2022simulated,
  title={Simulated bifurcation for higher-order cost functions},
  author={Kanao, Taro and Goto, Hayato},
  journal={Applied Physics Express},
  volume={16},
  number={1},
  pages={014501},
  year={2022},
  publisher={IOP Publishing}
}

@article{borders2019integer,
  title={Integer factorization using stochastic magnetic tunnel junctions},
  author={Borders, William A and Pervaiz, Ahmed Z and Fukami, Shunsuke and Camsari, Kerem Y and Ohno, Hideo and Datta, Supriyo},
  journal={Nature},
  volume={573},
  number={7774},
  pages={390--393},
  year={2019},
  publisher={Nature Publishing Group UK London}
}

@article{cai2020power,
  title={Power-efficient combinatorial optimization using intrinsic noise in memristor Hopfield neural networks},
  author={Cai, Fuxi and Kumar, Suhas and Van Vaerenbergh, Thomas and Sheng, Xia and Liu, Rui and Li, Can and Liu, Zhan and Foltin, Martin and Yu, Shimeng and Xia, Qiangfei and others},
  journal={Nature Electronics},
  volume={3},
  number={7},
  pages={409--418},
  year={2020},
  publisher={Nature Publishing Group UK London}
}

@article{aramon2019physics,
  title={Physics-inspired optimization for quadratic unconstrained problems using a digital annealer},
  author={Aramon, Maliheh and Rosenberg, Gili and Valiante, Elisabetta and Miyazawa, Toshiyuki and Tamura, Hirotaka and Katzgraber, Helmut G},
  journal={Frontiers in Physics},
  volume={7},
  pages={48},
  year={2019},
  publisher={Frontiers Media SA}
}

@article{mcmahon2016fully,
  title={A fully programmable 100-spin coherent Ising machine with all-to-all connections},
  author={McMahon, Peter L and Marandi, Alireza and Haribara, Yoshitaka and Hamerly, Ryan and Langrock, Carsten and Tamate, Shuhei and Inagaki, Takahiro and Takesue, Hiroki and Utsunomiya, Shoko and Aihara, Kazuyuki and others},
  journal={Science},
  volume={354},
  number={6312},
  pages={614--617},
  year={2016},
  publisher={American Association for the Advancement of Science}
}

@article{lucas2014ising,
  title={Ising formulations of many NP problems},
  author={Lucas, Andrew},
  journal={Frontiers in physics},
  volume={2},
  pages={5},
  year={2014},
  publisher={Frontiers Media SA}
}

@article{griset2024smartcharging,
  title={Software Suite for Benchmarking of Quantum Algorithms Applied to Two Typical Smart-Charging Optimization Problems},
  author={Griset, Rodolphe and Mikael, Joseph},
  year={2024}
}

@article{olariu2025set,
  title={A set packing model for the Partition Coloring Problem},
  author={Olariu, Emanuel Florentin and Fr{\u{a}}sinaru, Cristian},
  journal={Carpathian Journal of Mathematics},
  volume={41},
  number={2},
  pages={465--477},
  year={2025},
  publisher={JSTOR}
}

@article{cseker2019decomposition,
  title={A decomposition approach to solve the selective graph coloring problem in some perfect graph families},
  author={{\c{S}}eker, Oylum and Ekim, T{\i}naz and Ta{\c{s}}k{\i}n, Z Caner},
  journal={Networks},
  volume={73},
  number={2},
  pages={145--169},
  year={2019},
  publisher={Wiley Online Library}
}

@article{SEKER202167,
title = {An exact cutting plane algorithm to solve the selective graph coloring problem in perfect graphs},
journal = {European Journal of Operational Research},
volume = {291},
number = {1},
pages = {67-83},
year = {2021},
issn = {0377-2217},
author = {Oylum Şeker and Tınaz Ekim and Z. Caner Taşkın},
}

@inproceedings{fidanova2014ant,
  title={An ant algorithm for the partition graph coloring problem},
  author={Fidanova, Stefka and Pop, Petric{\u{a}} C},
  booktitle={International Conference on Numerical Methods and Applications},
  pages={78--84},
  year={2014},
  organization={Springer}
}

@article{furini2018exact,
  title={An exact algorithm for the partition coloring problem},
  author={Furini, Fabio and Malaguti, Enrico and Santini, Alberto},
  journal={Computers \& Operations Research},
  volume={92},
  pages={170--181},
  year={2018},
  publisher={Elsevier}
}

@article{kheiri2021constructing,
  title={Constructing operating theatre schedules using partitioned graph colouring techniques},
  author={Kheiri, Ahmed and Lewis, Rhyd and Thompson, Jonathan and Harper, Paul},
  journal={Health Systems},
  volume={10},
  number={4},
  pages={286--297},
  year={2021},
  publisher={Taylor \& Francis}
}

@article{zhu2020partition,
  title={Partition independent set and reduction-based approach for partition coloring problem},
  author={Zhu, Enqiang and Jiang, Fei and Liu, Chanjuan and Xu, Jin},
  journal={IEEE Transactions on Cybernetics},
  volume={52},
  number={6},
  pages={4960--4969},
  year={2020},
  publisher={IEEE}
}

@inproceedings{li2000partition,
  title={The partition coloring problem and its application to wavelength routing and assignment},
  author={Li, Guangzhi and Simha, Rahul},
  booktitle={Proceedings of the First Workshop on Optical Networks},
  volume={1},
  year={2000},
  organization={Dallas}
}

@article{cseker2022digital,
  title={Digital annealer for quadratic unconstrained binary optimization: a comparative performance analysis},
  author={{\c{S}}eker, Oylum and Tanoumand, Neda and Bodur, Merve},
  journal={Applied Soft Computing},
  volume={127},
  pages={109367},
  year={2022},
  publisher={Elsevier}
}

@article{zhao2024reinforcement,
  title={Reinforcement learning for electric vehicle charging scheduling: A systematic review},
  author={Zhao, Zhonghao and Lee, Carman KM and Yan, Xiaoyuan and Wang, Haonan},
  journal={Transportation Research Part E: Logistics and Transportation Review},
  volume={190},
  pages={103698},
  year={2024},
  publisher={Elsevier}
}

@article{dolgui2025scheduling,
  title={Scheduling electric vehicle regular charging tasks: A review of deterministic models},
  author={Dolgui, Alexandre and Kovalev, Sergey and Kovalyov, Mikhail Y},
  journal={European Journal of Operational Research},
  volume={325},
  number={2},
  pages={221--232},
  year={2025},
  publisher={Elsevier}
}

@article{zang2022column,
  title={A column generation tailored to electric vehicle routing problem with nonlinear battery depreciation},
  author={Zang, Yongsen and Wang, Meiqin and Qi, Mingyao},
  journal={Computers \& Operations Research},
  volume={137},
  pages={105527},
  year={2022},
  publisher={Elsevier}
}

@article{parmentier2023electric,
  title={Electric vehicle fleets: Scalable route and recharge scheduling through column generation},
  author={Parmentier, Axel and Martinelli, Rafael and Vidal, Thibaut},
  journal={Transportation Science},
  volume={57},
  number={3},
  pages={631--646},
  year={2023},
  publisher={Informs}
}

@article{chemudupaty2025optimizing,
  title={Optimizing trading of electric vehicle charging flexibility in the continuous intraday market under user and market uncertainties},
  author={Chemudupaty, Raviteja and Hornek, Timoth{\'e}e and Pavi{\'c}, Ivan and Menci, Sergio Potenciano},
  journal={Applied Energy},
  volume={381},
  pages={125103},
  year={2025},
  publisher={Elsevier}
}

@article{korkas2024distributed,
  title={Distributed and Multi-Agent Reinforcement Learning Framework for Optimal Electric Vehicle Charging Scheduling.},
  author={Korkas, Christos D and Tsaknakis, Christos D and Kapoutsis, Athanasios Ch and Kosmatopoulos, Elias},
  journal={Energies (19961073)},
  number={15},
  year={2024}
}

@article{wu2023electric,
  title={Electric vehicle charging scheduling considering infrastructure constraints},
  author={Wu, Ji and Su, Hao and Meng, Jinhao and Lin, Mingqiang},
  journal={Energy},
  volume={278},
  pages={127806},
  year={2023},
  publisher={Elsevier}
}

@article{xu2025impact,
  title={Impact and optimization of vehicle charging scheduling on regional clean energy power supply network management},
  author={Xu, Penghui and Wang, Xiaobo and Li, Zhichao},
  journal={Energy Informatics},
  volume={8},
  number={1},
  pages={13},
  year={2025},
  publisher={Springer}
}

@article{shen2025coordinating,
  title={Coordinating Day-Ahead and Intraday Scheduling for Bidirectional Charging of Fleet EVs},
  author={Shen, Shiwei and Haider, Syed Irtaza and Habeeb, Razan and Fitzek, Frank HP},
  journal={Automation},
  volume={6},
  number={4},
  pages={64},
  year={2025},
  publisher={Multidisciplinary Digital Publishing Institute}
}

@article{wang2024joint,
  title={Joint Modelling of Electric Vehicle Charging and Daily Activity Scheduling},
  author={Wang, Senlei and Pougala, Janody and Hillel, Tim},
  journal={Available at SSRN 5143388},
  year={2024}
}

\end{document}